\documentstyle[12pt,epsfig]{article}
\begin{document}
\topskip 2cm
\begin{titlepage}
\rightline{ DFUB 97-15 } 
\rightline{ hep-th/9711188 } 
\begin{center}
{\large\bf Differential Equations for Feynman Graph Amplitudes. } \\
\vspace{2.5cm}
{\large Ettore Remiddi } \\
\vspace{.5cm}
{\sl Dipartimento di Fisica, Universit\`a di Bologna, Italy \\ 
     INFN, Sezione di Bologna, Italy } \\ 
{\tt remiddi@bo.infn.it} 
\vspace{2.0cm}
\vfil
\begin{abstract}
It is by now well established that, by means of the integration by part 
identities \cite{ChetTka}, all the integrals occurring in the evaluation 
of a Feynman graph of given topology can be expressed in terms of a few 
independent master integrals. It is shown in this paper that the 
integration by part identities can be further used for obtaining a linear 
system of first order differential equations for the master integrals 
themselves. The equations can then be used for the numerical evaluation 
of the amplitudes as well as for investigating their analytic properties, 
such as the asymptotic and threshold behaviours and the corresponding 
expansions (and for analytic integration purposes, when possible). \par 
The new method is illustrated through its somewhat detailed 
application to the case of the one loop self-mass amplitude, by explicitly 
working out expansions and quadrature formulas, both in arbitrary 
continuous dimension \( n \) and in the \( n \to 4 \) limit. \par 
It is then shortly discussed which features of the new method are expected 
to work in the more general case of multi-point, multi-loop amplitudes. 
\end{abstract}
\end{center}
\scriptsize{ \noindent ------------------------------- 
} 

\end{titlepage}
\pagestyle{plain} \pagenumbering{arabic} 
\def\a{\alpha} 
\def\app{{\left(\frac{\alpha}{\pi}\right)}} 
\newcommand{\labbel}[1]{\label{#1}} 
\section{Introduction}           \par 
The usefulness of the integration by parts identities \cite{ChetTka} for 
reducing 
dramatically the number of independent integrals to be evaluated in the 
calculation of multi-loop Feynman graph amplitudes, expressing them in 
terms of a small set of independent integrals (often called master 
integrals), is by now well established (see for instance Ref. \cite{Crad96}). 
It is the purpose of this paper to point out that the same technique 
can be used for obtaining a linear system of first order differential 
equations in any of the external Mandelstam variables for the 
master integrals themselves \cite{Kotikov}. \par 
The new differential equations provide with a fairly complete 
information on the analytic properties of the integrals. That information 
can be conveniently used, for instance, for the direct numerical evaluation 
of the integrals themselves, or for the analytic expansions around particular 
values of the external scalar variables or of the internal masses -- and that 
by merely algebraic methods, without any attempt at carrying out the loop 
integrations occurring in the definition of the Feynman graph.
The equations can also be exploited for attempting their explicit analytic 
integration (an achievement which is however expected to be possible only 
in a limited number of particular cases). 
As required by the integration by part algorithm, all the loop integrals 
are defined in \( n \) continuous dimensions; the \( n \to 4 \) 
limit is also easily worked out within the new approach. 
\par 
To give a flavour of the features of the method, the almost elementary 
cases of 1-loop vacuum (0-point) and self-mass (2-point) amplitudes is 
discussed in some details (in those cases the system of equations 
reduces actually to a single equation). While it is claimed that the 
approach can be extended, without significant loss of effectiveness, 
to more general multi-loop and multi-point amplitudes, it is obvious 
that the 1-loop 2-point case is too simple to be really significant, so 
that  from this point of view this paper can be considered just an 
introduction to further work (which is in progress). 
In any case, it is hoped that the particular application of the new method 
presented in this paper might be of interest on its own, at least 
pedagogically. \par 
The paper is articulated in seven Sections. This is the first 
Section; Section 2 sets the notation and deals with the 1-loop vacuum 
amplitude; Section 3 establishes the differential equation for the 
1-loop self-mass amplitude, which is the main result of the paper; Section 4 
discusses a number of expansions which can be worked out from the equation; 
Section 5 recasts the equation as a quadrature formula; Section 6 is 
devoted to the \( n\to4 \) limit; Section 7, finally, contains an outlook 
to the possible extensions of the method to multi-point, multi-loop 
amplitudes. 
\section{The 1-loop vacuum amplitude. } \par 
\def\dnk{ \frac{d^nk}{(2\pi)^{n-2}} } 
We will use a Euclidean \( n \)-dimensional loop variable \( k \), with 
volume element defined as 
\begin{equation} \dnk \ ; 
\labbel{01} \end{equation} 
the Minkowski loop variables are recovered by the Wick rotation 
\[ \int d^nk = \int d^{n-1}k\; dk_n 
             = -i\left(\int d^{n-1}k\; dk_0 \right)_{\mathrm{Mink}} 
             = -i\left(\int d^nk \right)_{\mathrm{Mink}} \ .            \] 
The simplest 1-loop 0-point graph is shown in Fig.1; the associated 
amplitudes are 
\begin{equation} 
  T(n,m^2,-\a) = \int \dnk \; \frac{1}{(k^2+m^2)^\a} \ , 
\labbel{02} \end{equation} 
where \( \a \) is a positive integer. \par 
\vskip 3.0 truecm 
\epsfbox[0 20 20 40]{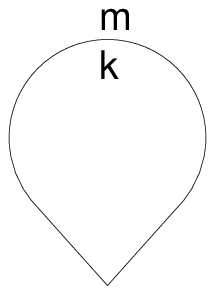} 
\centerline{Fig.1: The 1-loop vacuum graph.} 
\vspace{\baselineskip} 
An alternative dimensionless definition is 
\begin{equation} 
  T(n,(m/\mu)^2,-\a) = \mu^{2\alpha-n} \int \dnk \; \frac{1}{(k^2+m^2)^\a} \ , 
\labbel{02a} \end{equation} 
where \( \mu \) is a scale with the dimension of a mass. 
For simplicity, we will however use in the following Eq.(\ref{02}), 
corresponding to Eq.(\ref{02a}) at \( \mu = 1 \). 
The integration by parts identities are obtained starting from 
\[ \int \dnk \frac{\partial}{\partial k_\mu} 
             \left( k_\mu \; \frac{1}{(k^2+m^2)^\a} \right) = 0 \ , \] 
(the integral of the divergence vanishes for small enough \( n \)); 
carrying out the algebra, one obtains 
\begin{equation} 
  T(n,m^2,-\a) = \left( 1 - \frac{n}{2(\a-1)} \right) 
                \frac{1}{m^2} T(n,m^2,-\a+1) \ , 
\labbel{03a} \end{equation} 
or the equivalent equation 
\begin{equation} 
  T(n,m^2,-\a) = \frac{2\a}{2\a-n} m^2 T(n,m^2,-\a-1) \ . 
\labbel{03b} \end{equation} 
With Eq.(\ref{03a}) all the integrals with \( \a > 1 \) are expressed in 
terms of the ``master integral" corresponding to \( \a=1 \), 
\begin{equation} T(n,m^2) = T(n,m^2,-1) = \int \dnk \; \frac{1}{k^2+m^2} \ , 
\labbel{04} \end{equation} 
while the continuation of Eq.(\ref{03b}) to \( \a = 0 \) gives 
\[ T(n,m^2,0) = \int \dnk = 0 \] 
(the same applies to negative values of \( \a \) ). \par 
\( T(n,m^2) \) is divergent in the \( n \to 4 \) limit; by using twice 
Eq.(\ref{03a}) one gets 
\begin{equation} T(n,m^2) = \frac{8m^4}{(n-2)(n-4)}T(n,m^2,-3) \ , 
\labbel{05} \end{equation} 
which provides with a convenient analytic continuation in \( n \), as 
\( T(n,m^2,-3) \) is finite at \( n=4 \). According to the definition 
Eq.(\ref{02}), scaling \( k \) by \( m \) and then expressing the loop 
momentum in \( n \)-spherical coordinates, 
\( d^nk = m^n K^{n-1}dK d\Omega(n) \) one easily obtains 
\begin{equation} T(n,m^2,-3) = \frac{m^{n-6}\Omega(n)}{(2\pi)^{n-2}} 
         \int\limits_0^\infty \frac{K^{n-1}dK}{(K^2+1)^3} \ , 
\labbel{06} \end{equation} 
where \( \Omega(n) \) is the \( n \)-dimensional solid angle. At 
\( n=4 \), a simple explicit calculation gives 
\[ T(4,m^2,-3) = \frac{1}{8m^2} \ ; \] 
one can then write 
\begin{equation} 
   T(n,m^2) = \frac{m^{n-2}}{(n-2)(n-4)} C(n) \ , 
\labbel{07} \end{equation} 
where \( C(n) \) is a suitable dimensionless function of \( n \) , defined as 
\begin{equation} C(n) = 8  T(n,1,-3) \ , 
\labbel{08} \end{equation} 
with the known limiting value at \( n=4 \) 
\begin{equation} C(4) = 1 \ . \labbel{09} \end{equation} 
In most applications \( C(n) \) can be kept as an overall factor; its 
expansion in \( n \) around \( n = 4 \), which can be obtained from 
Eq.s(\ref{08},\ref{06}), is actually not needed if \( C(n) \) multiplies a 
finite expression in which the \( 1/(n-4) \) singularities have cancelled out. 
\par 
It is further to be noted that for \( n>2 \) Eq.(\ref{07}) implies 
\begin{equation} T(n,0) = 0 \ . \labbel{09a} \end{equation} 
\par 
For convenience of later use, let us also look at the 1-loop, 2-denominator 
vacuum graph of Fig.2. 
\\ \vbox{ 
\vskip 2.5 truecm 
\epsfbox[0 20 20 40]{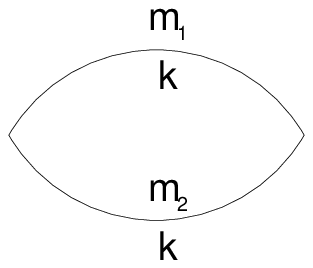} 
\centerline{Fig.2: The 1-loop, 2-denominator vacuum graph.} 
\vspace{\baselineskip} 
\noindent 
      } \\ 
Considering for simplicity only the case in which both denominators are 
raised to the first power, by partial fractioning one finds 
\begin{equation} 
   \int \dnk \; \frac{1}{(k^2+m_1^2)(k^2+m_2^2)} = 
        \frac{1}{m_2^2-m_1^2} \left( T(n,m_1^2) - T(n,m_2^2) \right)\ , 
\labbel{010} \end{equation} 
showing that the amplitudes associated to the graph of Fig.2 are just 
algebraic combinations of the amplitudes \( T(n,m^2,\a) \) associated 
to Fig.1. 
\section{The 1-loop self-mass amplitude. } \par 
The 1-loop 2-point (self-mass) graph is shown in Fig.3. 
\vskip 2.5 truecm 
\epsfbox[0 20 20 40]{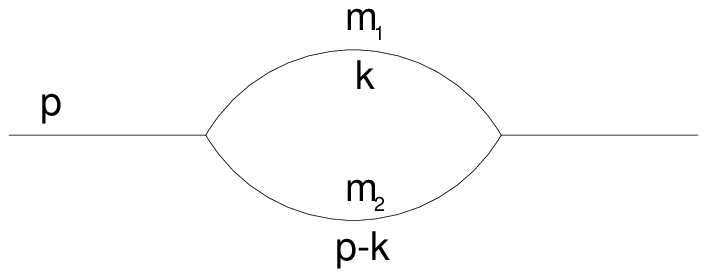} 
\centerline{Fig.3: The 1-loop self-mass graph.} 
\vspace{\baselineskip} 
\noindent 
\newcommand{\Sg}[2]{S(n,m_1^2,m_2^2,#1,#2,p^2)} 
\newcommand{\Sm}{S(n,m_1^2,m_2^2,p^2)} 
The associated scalar integrals are 
\begin{equation} 
  \Sg{-\a_1}{-\a_2} = \int \dnk \; \frac{1} 
                   { (k^2+m_1^2)^{\a_1} ((p-k)^2+m_2^2)^{\a_2} } \ , 
\labbel{10} \end{equation} 
where \( \a_1,\a_2 \) are positive integers and 
the external vector \( p \) is also Euclidean ( \( p^2>0 \) 
if \( p \) is spacelike; the timelike region is to be recovered by means 
of the usual analytic continuation). The integration by parts identities 
are derived from the equations 
\begin{equation} 
 \int \dnk \frac{\partial}{\partial k_\mu} \left( v_\mu \; \frac{1} 
         { (k^2+m_1^2)^{\a_1} ((p-k)^2+m_2^2)^{\a_2} } \right) = 0 \ , 
\labbel{10a} \end{equation} 
where the vector \( v_\mu \) can be either \( p_\mu \) or \( k_\mu \), 
so that there are 2 identities for each choice of the pair \( \a_1,\a_2 \). 
Writing down explicitly the identities is elementary; 
as a consequence of the identities, it turns out that all the integrals 
defined in Eq.(\ref{10}) with \( \a_1 \) or \( \a_2 \) larger than 1 can be 
expressed in terms of a single master integral corresponding to 
\( \a_1 = \a_2 = 1 \), namely 
\begin{equation} 
  \Sm = \int \dnk \; \frac{1} { (k^2+m_1^2) ((p-k)^2+m_2^2) } \ . 
\labbel{11} \end{equation} 
Indeed, by writing explicitly the two identities corresponding to 
\( \a_1=\a_2=1 \) in Eq.(\ref{10a}), after some elementary algebra one finds 
\begin{eqnarray} 
    \Sg{-1}{-2} = {\kern-20pt} && 
          \frac{1} {[p^2+(m_1+m_2)^2][p^2-(m_1+m_2)^2]} \nonumber \\ 
   && \cdot \left[ {\kern-15pt} \phantom{\frac{p^2}{m^2} } 
      (n-3)(m_1^2-m_2^2-p^2) \Sm + (n-2) T(n,m_1^2) \right.   \nonumber \\ 
   && \left. {\kern+5pt} 
      - (n-2) \frac{p^2+m_1^2+m_2^2}{2m_2^2} T(n,m_2^2) \right] \ . 
\labbel{12} \end{eqnarray} 
The similar formula for \( \Sg{-2}{-1} \) is then immediately obtained 
by exchanging \( m_1 \) with \( m_2 \) in Eq.(\ref{12}); 
by further using the relation 
\begin{equation} \Sg{-(\a_1+1)}{-\a_2} = - \frac{1}{\a_1} 
              \frac{\partial}{\partial m_1^2} \Sg{-\a_1}{-\a_2} \ , 
\labbel{13} \end{equation} 
which follows at once from the definition Eq.(\ref{10}), one obtains 
the corresponding equations for higher values of \( \a_1,\a_2 \). \par 
Summing up: all the integrals \( \Sg{-\a_1}{-\a_2} \), Eq.(\ref{10}), 
can be expressed as a combination of the self-mass master integral \( \Sm \) 
and the vacuum master integral \( T(n,m^2) \), defined in Eq.(\ref{11}) and 
Eq.(\ref{04}), times the ratio of suitable polynomials in \( p^2 \), 
\( m_1^2 \) and \( m_2^2 \) --- a well established and already known result. 
\par 
We consider now the \( p^2 \)-derivative of \( \Sm \). We can obviously 
write 
\begin{equation} p^2 \frac{\partial}{\partial p^2} \Sm = 
              \frac{1}{2}\ p_\mu \frac{\partial}{\partial p_\mu} \Sm \ . 
\labbel{14} \end{equation} 
By using the definition of \( \Sm \), Eq.(\ref{11}), and 
then carrying out explicitly the \mbox{\( p_\mu \)-derivatives,} one obtains 
\[ p_\mu \frac{\partial}{\partial p_\mu} \Sm = 
   -2 \int \dnk \; \frac{1}{k^2+m_1^2} \; 
                   \frac{p^2-p\cdot k}{((p-k)^2+m_2^2)^2} \ ;     \] 
with the obvious identity 
\[ p\cdot k = \frac{1}{2} \left[ (k^2+m_1^2) - ((p-k)^2+m_2^2) 
                                     +p^2-m_1^2+m_2^2 \right] \ , \] 
the previous equation becomes 
\begin{eqnarray} p_\mu \frac{\partial}{\partial p_\mu} \Sm = 
     && {\kern -15 pt} - (p^2+m_1^2-m_2^2) \Sg{-1}{-2} \nonumber \\ 
     && {\kern -10 pt} - \Sm + T(n,m_2^2,-2) \ ;    \nonumber 
\end{eqnarray} 
by using Eq.(\ref{03a}) and Eq.(\ref{12}), after minor rearrangements 
Eq.(\ref{14}) can be written as 
\begin{eqnarray} 
\frac{\partial}{\partial p^2} \Sm = && {\kern -15pt} 
       - \frac{n-2}{2p^2} \Sm            \nonumber \\ 
   && {\kern -120pt} + (n-3) 
     \frac{p^2+m_1^2+m_2^2}{[p^2+(m_1+m_2)^2][p^2+(m_1-m_2)^2]} \Sm 
                                                      \nonumber \\ 
   && {\kern -120pt} - (n-2) 
      \frac{(p^2+m_1^2-m_2^2)T(n,m_1^2) + (p^2-m_1^2+m_2^2)T(n,m_2^2)} 
           {2p^2[p^2+(m_1+m_2)^2][p^2+(m_1-m_2)^2]} \ , 
\labbel{15} \end{eqnarray} \par 
That same equation can be derived, alternatively, starting from the 
definition of \( \Sm \) Eq.(\ref{11}) and replacing the loop variable 
\( k \) by \(k/\lambda \), obtaining 
\[ \Sm = \lambda^{4-n} S(n,\lambda^2 m_1^2, 
         \lambda^2 m_2^2,\lambda^2 p^2) \ ; \] 
by acting on both sides with \( \lambda^2 \frac{ \partial } 
{\partial \lambda^2 } \) one gets the familiar scaling equation 
\[ \left(  p^2 \frac{\partial}{\partial p^2} 
  + m_1^2 \frac{\partial}{\partial m_1^2} 
  + m_2^2 \frac{\partial}{\partial m_2^2} 
  + \frac{4-n}{2} \right) \Sm = 0 \ , \] 
which according to Eq.(\ref{13}) can also be written as 
\begin{eqnarray*} 
  p^2 \frac{\partial}{\partial p^2} \Sm = & & \frac{n-4}{2} \Sm \\ 
         &+& \Sg{-2}{-1} + \Sg{-1}{-2} \ ; 
\end{eqnarray*} 
by substituting Eq.(\ref{12}) and the similar formula for \( \Sg{-2}{-1} \), 
Eq.(\ref{15}) is recovered. \par 
Eq.(\ref{15}) is the central result of this paper. It is an inhomogeneous, 
first-order differential equation in the external variable \( p^2 \) 
for the master integral \( \Sm \); the coefficients are ratios of 
polynomials in \( p^2 \) and the masses, the inhomogeneous term is 
known, consisting of the simpler vacuum amplitudes \( T(n,m^2) \) 
discussed at length in the previous section. As Eq.(\ref{15}) is a first 
order equation in \( p^2 \), it determines \( \Sm \) once a single 
``initial value" for some value of the variable \( p^2 \) is provided; 
given that initial value, the numerical integration of the equation 
up to any desired value of \( p^2 \) can be easily carried out 
along any path in the complex \( p^2 \)-plane --- and that for arbitrary 
continuous dimension \( n \), the \( n \to 4 \) limit being just a 
particular case (see below). \par 
Eq.(\ref{15}) can also be used for studying the analytic properties of 
\( \Sm \), its value at specific values of \( p^2 \) or, more generally, 
its expansion around those values. Eq.(\ref{15}) is particularly well suited 
for the discussion of the properties of \( \Sm \) at the potentially 
singular points of the equation, namely 
\( p^2 = 0 \), \hbox{\( p^2 = -(m_1-m_2)^2 \)}, 
\hbox{\( p^2 = -(m_1+m_2)^2 \)} and \( p^2 \to \infty \); Eq.(\ref{15}), 
in addition, can also be recast 
in the form of a quadrature formula which can then be used for the actual
analytical integration in the \( n \to 4 \) limit. 
\section{Expansions of the 1-loop self-mass amplitude. } \par 
Let us start to discuss the point \( p^2 = 0 \); to that aim, Eq.(\ref{15}) 
can be rewritten in the fully equivalent way \\ \vbox{ 
\begin{eqnarray} 
\frac{\partial}{\partial p^2} \Sm = && {\kern -15pt} 
       - \frac{n-2}{2p^2} \left( \Sm + 
       \frac{T(n,m_1^2) - T(n,m_2^2)}{m_1^2-m_2^2} \right) 
                             \nonumber \\ 
   && {\kern -120pt} + \frac{n-3}{2}\left( \frac{1}{p^2+(m_1+m_2)^2} 
            + \frac{1}{p^2+(m_1-m_2)^2} \right) \Sm \nonumber \\ 
   && {\kern -120pt} + \frac{n-2}{4m_1(m_1^2-m_2^2)} \left( 
       \frac{m_1-m_2}{p^2+(m_1+m_2)^2} + \frac{m_1+m_2}{p^2+(m_1-m_2)^2} 
       \right) T(n,m_1^2) \nonumber \\ 
   && {\kern -120pt} + \frac{n-2}{4m_2(m_1^2-m_2^2)} \left( 
       \frac{m_1-m_2}{p^2+(m_1+m_2)^2} - \frac{m_1+m_2}{p^2+(m_1-m_2)^2} 
       \right) T(n,m_2^2) \ . 
\labbel{16} \end{eqnarray} } \par 
 By inspection of 
the definition Eq.(\ref{11}) (but without any explicit integration on 
the loop momentum \( k \)) \( \Sm \) is seen to be regular at 
\( p^2 = 0 \), {\it i.e.} the function and its derivatives are finite 
at that point. The apparent singularity of Eq.(\ref{16}) at \( p^2 = 0 \) 
must therefore disappear: that amounts to the condition 
\begin{equation} 
  S(n,m_1^2,m_2^2,0) = - \frac{T(n,m_1^2)-T(n,m_2^2)}{m_1^2-m_2^2} \ , 
\labbel{17} \end{equation} 
where \( T(n,m^2) \) is known, as already remarked. 
In other words, the equation itself gives the explicit value of 
\( S(n,m_1^2,m_2^2,0) \), once the information of the regularity at 
\( p^2 = 0 \) is provided. That same result, Eq.(\ref{17}) may be also 
obtained by noting that at \( p_\mu = 0 \) (implying \( p^2 = 0 \) ) 
Eq.(\ref{11}) goes into Eq.(\ref{010}). \par 
Knowing that \( \Sm \) is regular at \( p^2 = 0 \), we can also expand 
\( \Sm \) in \( p^2 \) around that point 
\begin{equation} 
  \Sm = \sum \limits_{k=0}^{\infty} S_k(n,m_1^2,m_2^2) (p^2)^k \ , 
\labbel{18} \end{equation} 
where \( S_0(n,m_1^2,m_2^2) = S(n,m_1^2,m_2^2,0) \) is of course given 
by Eq.(\ref{17}). By inserting the expansion Eq.(\ref{18}) into Eq.(\ref{16}) 
the coefficients \( S_k(n,m_1^2,m_2^2) \) are recursively determined; the 
first explicit values are 
\begin{eqnarray}
  S_1(n,m_1^2,m_2^2) &=& \frac{1}{n(m_1^2-m_2^2)^3} \left[ 
     - (n-4)(m_1^2+m_2^2)\left( T(n,m_1^2) - T(n,m_2^2) \right) 
      \right. \nonumber \\ && {\kern +65pt} \left. 
     +2(n-2)\left( m_2^2T(n,m_1^2) - m_1^2T(n,m_2^2) \right) \right] \ , 
              \nonumber \\ 
  S_2(n,m_1^2,m_2^2) &=& \frac{1}{n(n+2)(m_1^2-m_2^2)^5} \nonumber \\ 
    && \left[ (n-4)\left( 6 (m_1^2+m_2^2)^2 - n (m_1^2-m_2^2)^2 \right) 
           \left( T(n,m_1^2) - T(n,m_2^2) \right) \right. \nonumber \\ 
    && \left. - 12 (n-2) (m_1^2+m_2^2) 
       \left( m_2^2T(n,m_1^2) - m_1^2T(n,m_2^2) \right) \right] \ . 
\labbel{19} \end{eqnarray} 
Note that the knowledge of the coefficients of the expansion around 
\( p^2=0 \) allows also the accurate evaluation of the {\it r.h.s} of 
Eq.(\ref{16}) for small values of \( p^2 \) when trying its numerical 
integration starting from \( p^2=0 \). \par 
A similar approach can be in principle worked out also at the point 
\( p^2 = -(m_1-m_2)^2 \). It is known, from the conventional study of the 
analytic properties of \( \Sm \) (which will not be repeated here), that 
\( \Sm \) is also regular at that particular value of \( p^2 \) (sometimes 
referred to as pseudo-threshold). Imposing the regularity at that 
point Eq.(\ref{16}) gives 
\begin{equation} 
  S(n,m_1^2,m_2^2,-(m_1-m_2)^2) = - \frac{n-2}{n-3}\; 
               \frac{m_2T(n,m_1^2)-m_1T(n,m_2^2)}{2m_1m_2(m_1-m_2)} \ . 
\labbel{19a} \end{equation} 
It is a somewhat striking feature of this approach 
that the condition of regularity at \( p^2 = 0 \), determining completely 
\( \Sm \) through Eq.(\ref{16}), which is a first order differential 
equation, implies also the 
regularity at \( p^2 = -(m_1-m_2)^2 \), a result which we were however 
unable to derive directly from Eq.(\ref{16}) only, {\it i.e.} without 
explicit reference to the definition Eq.(\ref{11}). \par 
\newcommand{\Sy}{\Phi(n,m_1^2,m_2^2,y)} 
For discussing the \( p^2 \to \infty \) limit, let us switch to the usual 
inverse variable \( y = 1/p^2 \), introducing the function 
\begin{equation} \Sy = S\left(n,m_1^2,m_2^2,\frac{1}{y}\right) \ . 
\labbel{20} \end{equation} 
In terms of the new variable \( y \), Eq.(\ref{16}) becomes \\ 
\vbox{ \begin{eqnarray} 
\frac{\partial}{\partial y} \Sy= && 
       - \frac{n-4}{2y} \Sy     {\kern -10pt} \nonumber \\ 
   && {\kern -30pt} + \frac{n-3}{2}\left( \frac{1}{y+\frac{1}{(m_1+m_2)^2}} 
            + \frac{1}{y+\frac{1}{(m_1-m_2)^2}} \right) \Sy \nonumber \\ 
   && {\kern -30pt} + \frac{n-2}{4m_1(m_1^2-m_2^2)} 
  \left( \frac{m_1-m_2}{y+\frac{1}{(m_1+m_2)^2}} 
       + \frac{m_1+m_2}{y+\frac{1}{(m_1-m_2)^2}} \right) T(n,m_1^2) 
                                                            \nonumber \\ 
   && {\kern -30pt} + \frac{n-2}{4m_2(m_1^2-m_2^2)} 
  \left( \frac{m_1-m_2}{y+\frac{1}{(m_1+m_2)^2}} 
       - \frac{m_1+m_2}{y+\frac{1}{(m_1-m_2)^2}} \right) T(n,m_2^2) \ . 
\labbel{21} \end{eqnarray} } 
The equation exhibits at \( y = 0 \) the singular behaviour 
\( y^{-(n-4)/2} \), so that the function must be expanded for 
small \( y \) as the sum of two series corresponding to the singular and 
regular parts 
\begin{equation} 
  \Sy = y^{-(n-4)/2} \sum \limits_{k=0}^{\infty} 
                       \Phi^{(s)}_k(n,m_1^2,m_2^2) y^k 
    + \sum \limits_{k=0}^{\infty} \Phi^{(r)}_k(n,m_1^2,m_2^2) y^k \ . 
\labbel{22} \end{equation} 
According to Eq.s(\ref{20},\ref{22}), the asymptotic expansion of \( \Sm \) 
for large \( p^2 \) is also given by 
\begin{equation} 
  \Sm = (p^2)^{(n-4)/2} \sum \limits_{k=0}^{\infty} 
   \Phi^{(s)}_k(n,m_1^2,m_2^2) \frac{1}{(p^2)^k}  + \sum 
   \limits_{k=0}^{\infty} \Phi^{(r)}_k(n,m_1^2,m_2^2) \frac{1}{(p^2)^k} \ . 
\labbel{23} \end{equation} 
By inserting the expansion Eq.(\ref{22}) into Eq.(\ref{21}) -- or more 
directly the expansion Eq.(\ref{23}) into Eq.(\ref{16}), disregarding 
from now on the variable \( y \) -- and imposing the equality of the 
coefficients of the same powers of the expansion parameter one obtains 
\begin{eqnarray} 
 \Phi^{(r)}_0(n,m_1^2,m_2^2) &=& 0 \ , \nonumber \\ 
 \Phi^{(r)}_1(n,m_1^2,m_2^2) &=& T(n,m_1^2) + T(n,m_2^2) \ , \nonumber \\ 
 \Phi^{(r)}_2(n,m_1^2,m_2^2) &=& 
     \left( \frac{n-4}{n} m_1^2 - m_2^2 \right) T(n,m_1^2) 
   + \left( \frac{n-4}{n} m_2^2 - m_1^2 \right) T(n,m_2^2) \ , 
\labbel{24} \end{eqnarray}
and 
\begin{eqnarray} 
 \Phi^{(s)}_1(n,m_1^2,m_2^2) &=& (n-3)( m_1^2 + m_2^2 ) 
             \Phi^{(s)}_0(n,m_1^2,m_2^2) \ , \nonumber \\ 
 \Phi^{(s)}_2(n,m_1^2,m_2^2) &=& 
     \frac{1}{2} (n-3) \left[(n-4)( m_1^4 + m_2^4 ) + 2 (n-6) m_1^2 m_2^2 
       \right] \Phi^{(s)}_0(n,m_1^2,m_2^2) \ . 
\labbel{25} \end{eqnarray} 
Note that Eq.(\ref{16}), by itself, is unable to fix completely all 
the coefficients of the expansion; the overall normalization of the 
singular part, \( \Phi^{(s)}_0(n,m_1^2,m_2^2) \) is in fact as yet not 
determined. But some more information on \( \Phi^{(s)}_0(n,m_1^2,m_2^2) \) 
can be obtained from Eq.(\ref{12}), 
by writing it, according to Eq.(\ref{13}), in the form 
\begin{eqnarray} 
    \frac{\partial}{\partial m_2^2} \Sm = {\kern-15pt} && - 
          \frac{1} {[p^2+(m_1+m_2)^2][p^2-(m_1+m_2)^2]} \nonumber \\ 
   && {\kern-35pt} \left[ \phantom{\frac{p^2}{m_1^2}}{\kern-15pt} 
      (n-3)(m_1^2-m_2^2-p^2) \Sm + (n-2) T(n,m_1^2) \right.   \nonumber \\ 
   && \left. {\kern-25pt} 
      - (n-2) \frac{p^2+m_1^2+m_2^2}{2m_2^2} T(n,m_2^2) \right] \ ; 
\labbel{26} \end{eqnarray} 
by inserting in it the large-\( p^2 \) expansion Eq.(\ref{23}) and then 
equating the coefficients of the same powers of \( p^2 \) we obtain 
the additional equation 
\begin{equation} 
  \frac{\partial}{\partial m_2^2} \Phi^{(s)}_0(n,m_1^2,m_2^2) = 0 \ ; 
\labbel{27} \end{equation} 
as the same is true for the derivative with respect to \( m_1^2 \), we 
can write as a conclusion 
\begin{equation} 
    \Phi^{(s)}_0(n,m_1^2,m_2^2) = \Phi^{(s)}_0(n) \ , 
\labbel{28} \end{equation} 
expressing the fact that the the still unknown leading coefficient 
\( \Phi^{(s)}_0(n) \) is independent of the masses, but depends only on 
the dimension \( n \). \par 
Due to the condition \( \Phi^{(r)}_0(n,m_1^2,m_2^2) = 0 \), 
Eq.(\ref{24}) and of the expansion Eq.(\ref{23}), for \( n > 2 \) 
Eq.(\ref{28}) implies 
\begin{equation} 
 \lim_{p^2\to \infty} (p^2)^{-(n-4)/2} \Sm = \Phi^{(s)}_0(n) \ , 
\labbel{29} \end{equation} 
while for smaller values of \( n \) the dominant term of the expansion 
is \( \Phi^{(r)}_1(n,m_1^2,m_2^2)/(p^2) \). \par 
The point \( p^2 = -(m_1+m_2)^2 \), finally, corresponds to the usual 
threshold and is therefore another singular point for Eq.(\ref{16}) and 
\( \Sm \). It can be discussed, when required, along the lines sketched 
for \( p^2 \to \infty \). In analogy to Eq.(\ref{23}) one can write the 
expansion around \( p^2 = -(m_1+m_2)^2 \) as the sum of two series,  
\begin{eqnarray} 
  \Sm &=& (p^2+(m_1+m_2)^2)^{(n-3)/2} \sum \limits_{k=0}^{\infty} 
   \Psi^{(s)}_k(n,m_1^2,m_2^2) (p^2+(m_1+m_2)^2)^k \nonumber \\ 
   &+& \sum \limits_{k=0}^{\infty} \Psi^{(r)}_k(n,m_1^2,m_2^2) 
    (p^2+(m_1+m_2)^2)^k \ ; 
\labbel{29a} \end{eqnarray} 
by inserting the expansion in Eq.(\ref{16}), it turns out that the 
coefficients \( \Psi^{(r)}_k(n,m_1^2,m_2^2) \) are completely fixed, while 
the \( \Psi^{(s)}_k(n,m_1^2,m_2^2) \) are all proportional to a same constant, 
which can be taken to be \( \Psi^{(s)}_0(n,m_1^2,m_2^2) \) and is not 
determined by Eq.(\ref{16}) alone. The first values of the coefficients 
are 
\begin{eqnarray} 
  \Psi^{(r)}_0(n,m_1^2,m_2^2) &=& -\frac{(n-3)(n-2)}{2m_1m_2(m_1+m_2)} 
      \left( {m_2}T(n,m_1^2) + {m_1}T(n,m_2^2) \right)    \nonumber \\ 
  \Psi^{(s)}_1(n,m_1^2,m_2^2) &=& \left( \frac{n-2}{2(m_1+m_2)^2} 
           - \frac{n-3}{8m_1m_2} \right) \Psi^{(s)}_0(n,m_1^2,m_2^2) \ . 
\labbel{29b} \end{eqnarray} 
\par 
As already done for the large \( p^2 \) expansion, one can insert the 
expansion Eq.(\ref{29a}) and the explicit values Eq.(\ref{29b}) in 
Eq.(\ref{26}), obtaining the equation 
\begin{equation} 
  \frac{\partial}{\partial m_2} \Psi^{(s)}_0(n,m_1^2,m_2^2) = 
      \left( \frac{n-3}{2m_2} - \frac{n-2}{m_1+m_2} \right) 
                                \Psi^{(s)}_0(n,m_1^2,m_2^2) \ . 
\labbel{29c} \end{equation} 
Its solution, symmetrized in \( m_1, m_2 \), is 
\begin{equation} 
  \Psi^{(s)}_0(n,m_1^2,m_2^2) = \frac{(m_1m_2)^{(n-3)/2}}{(m_1+m_2)^{n-2}} 
                                \Psi^{(s)}_0(n) \ , 
\labbel{29d} \end{equation} 
where the still unknown coefficient \( \Psi^{(s)}_0(n) \) depends only 
on \( n \). 
\section{The quadrature formula. } \par 
\newcommand{\sm}{\sigma(n,m_1^2,m_2^2,p^2)} 
The 1-loop self-mass case is simple enough to allow to recast the differential 
equation in the form of a quadrature formula. By putting 
\begin{eqnarray} 
 \Sm = &&(p^2)^{-(n-2)/2} \left[\left(p^2+(m_1+m_2)^2\right) 
            \left(p^2+(m_1-m_2)^2\right) \right]^{(n-3)/2} \nonumber \\ 
       && \cdot \sm \ , 
\labbel{30} \end{eqnarray} 
Eq.(\ref{15}) becomes 
\begin{eqnarray} 
  \frac{\partial}{\partial p^2} \sm = && {\kern-10pt} 
     - \frac{1}{2} (p^2)^{(n-4)/2} 
       \left[\left(p^2+(m_1+m_2)^2\right) 
             \left(p^2+(m_1-m_2)^2\right) \right]^{-(n-1)/2} \nonumber \\ 
    && {\kern-50pt} (n-2) 
       \left[ (p^2+m_1^2-m_2^2)T(n,m_1^2) + (p^2-m_1^2+m_2^2)T(n,m_2^2) 
       \right] \ . 
\labbel{31} \end{eqnarray} 
The above equation is equivalent to the quadrature formula 
\begin{eqnarray} 
  \sm &=& \sigma(n,m_1^2,m_2^2,\overline{p}^2) \nonumber \\ 
   && {\kern-70pt} - \frac{n-2}{2} \int\limits_{\overline{p}^2}^{p^2} dx 
         \ x^{(n-4)/2} \left[\left(x+(m_1+m_2)^2\right) 
             \left(x+(m_1-m_2)^2\right) \right]^{-(n-1)/2} \nonumber \\ 
   && {\kern-20pt} \cdot \left[ (x+m_1^2-m_2^2)T(n,m_1^2) 
                              + (x-m_1^2+m_2^2)T(n,m_2^2) \right] \ , 
\labbel{32} \end{eqnarray} 
where \( \overline{p}^2 \) is some suitable value of \( p^2 \). As 
\( \Sm \) is regular at \( p^2=0 \), for \( n>2 \) Eq.(\ref{30}) implies 
\( \sigma(n,m_1^2,m_2^2,0) = 0 \), so that \( \overline{p}^2 = 0 \) 
is a natural choice. \par 
For \( m_2 = 0 \), by using Eq.(\ref{09a}), one has 
\begin{equation} 
  \sigma(n,m^2,0,p^2) = - \frac{n-2}{2} T(n,m^2) \int\limits_{0}^{p^2} 
          dx \; x^{(n-4)/2} (x+m^2)^{-(n-2)} \ ; 
\labbel{33} \end{equation} 
which becomes, by using Eq.(\ref{07}) and scaling \( x \) by \( m^2 \) 
within the integrand, 
\begin{equation} 
  \sigma(n,m^2,0,p^2) = - \frac{C(n)}{2(n-4)} \int\limits_{0}^{p^2/m^2} 
          dx \; x^{(n-4)/2} (x+1)^{-(n-2)} \ . 
\labbel{34} \end{equation} 
By comparing Eq.(\ref{29}), Eq.(\ref{30}) and Eq.(\ref{34}) a closed 
expression for \( \Phi^{(s)}_0(n) \) is obtained 
\begin{equation} 
  \Phi^{(s)}_0(n) = - \frac{C(n)}{2(n-4)} \int\limits_{0}^{\infty} 
          dx \; x^{(n-4)/2} (x+1)^{-(n-2)} \ . 
\labbel{35} \end{equation} 
\section{The \( n \to 4 \) limit. } \par 
\newcommand{\Sbm}{\overline{S}(m_1^2,m_2^2,p^2)} 
\newcommand{\Sb}{\overline{S}(4,m_1^2,m_2^2,p^2)} 
It is known that in the \( n \to 4 \) limit 1-loop amplitudes are singular 
at most as \( 1/(n-4) \); from Eq.(\ref{07}) one has indeed, for \( n \to 4 \), 
\begin{equation} 
      T(n,m^2) = m^2 \left[ \frac{C(n)}{2(n-4)} 
   + \frac{1}{2} \ln{m} - \frac{1}{4} + { \cal O } (n-4) \right] , 
\labbel{39} \end{equation} 
where \( C(n) \) is given by Eq.s(\ref{08},\ref{09}). If \( \ln{m} \) is 
perceived as disturbing, replace Eq.(\ref{02}) with Eq.(\ref{02a}) and 
\( \ln{m} \) goes into \( \ln{(m/\mu)} \). \par 
Accordingly, \( \Sm \) can be expanded in the \( n \to 4 \) limit as 
\begin{equation} 
   \Sm = \frac{C(n)}{n-4} S_0(m_1^2,m_2^2,p^2) + S_1(m_1^2,m_2^2,p^2) \ . 
\labbel{39a} \end{equation} 
Eq.(\ref{15}) is an identity in \( n \); therefore, it implies a set 
of independent equations for the various terms of the expansion in 
\( (n-4) \). The part singular in \( (n-4) \), {\it i.e.} the coefficient 
of \( C(n)/(n-4) \), gives 
\begin{eqnarray} 
  \frac{\partial}{\partial p^2} S_0(m_1^2,m_2^2,p^2) = 
    - && {\kern-15pt} \left( \frac{1}{p^2} - \frac{1}{2(p^2+(m_1+m_2)^2)} 
                  - \frac{1}{2(p^2+(m_1-m_2)^2)} \right) \nonumber \\ 
      && {\kern-12pt} 
         \cdot \left( S_0(m_1^2,m_2^2,p^2) + \frac{1}{2} \right) \ . 
\labbel{39b} \end{eqnarray} 
On the other hand, we can expand \( S(n,m_1^2,m_2^2,0) \), Eq.(\ref{17}); 
that gives, in the notation of Eq.(\ref{39a}), 
\begin{equation} S_0(m_1^2,m_2^2,0) = - \frac{1}{2} 
\labbel{39c} \end{equation} 
and 
\begin{equation} S_1(m_1^2,m_2^2,0) = + \frac{1}{4} 
    - \frac{ m_1^2\ln{m_1} - m_2^2\ln{m_2} }{2(m_1^2-m_2^2)} \ . 
\labbel{39d} \end{equation} 
(Again, \( \ln{m_1} \) becomes \( \ln{(m_1/\mu)} \) etc. if the 
{\it r.h.s.} of Eq.(\ref{10}) is multiplied by \( \mu^{4-n} \)).  
\par 
With the input Eq.(\ref{39c}), Eq.(\ref{39b}) gives, for any \( p^2 \), 
\begin{equation} S_0(m_1^2,m_2^2,p^2) = - \frac{1}{2} \ , 
\labbel{39e} \end{equation} 
independent of \( p^2 \) and of the masses. \par 
For discussing the finite part of Eq.(\ref{15}) in the \( n \to 4 \) limit, 
it can be convenient to write \( S_1(m_1^2,m_2^2,p^2) \) as 
\begin{equation} S_1(m_1^2,m_2^2,p^2) = S_1(m_1^2,m_2^2,0) + \Sbm \ , 
\labbel{40} \end{equation} 
where \( S_1(m_1^2,m_2^2,0) \) is known from Eq.(\ref{39d}). 
\newcommand{\verso}[1]{ {\; \buildrel {n \to #1} \over{\longrightarrow}}\; } 
Note that Eq.(\ref{40}), due to Eq.(\ref{39e}),  amounts to 
\begin{equation} 
  \Sbm = \lim_{n\to4} \left( \Sm - S(n,m_1^2,m_2^2,0) \right) \ , 
\labbel{41} \end{equation} 
or, in equivalent way, 
\begin{equation} 
     \Sm \verso{4} S(n,m_1^2,m_2^2,0) + \Sbm \ . 
\labbel{41a} \end{equation} 
From Eq.(\ref{15}) -- or, rather, from its equivalent form Eq.(\ref{16}) -- 
one then obtains for \( \Sbm \) the equation 
\begin{eqnarray} 
  \frac{\partial}{\partial p^2} \Sbm = && - \frac{1}{p^2}\ \Sbm \nonumber \\ 
   && {\kern-120pt} + \frac{1}{2}\left( \frac{1}{p^2+(m_1+m_2)^2} 
            + \frac{1}{p^2+(m_1-m_2)^2} \right) \Sbm \nonumber \\ 
   && {\kern-120pt} + \frac{m_1^2m_2^2}{m_1^2-m_2^2} 
      \ln{\left( \frac{m_1}{m_2} \right) } 
      \ \frac{1} {[p^2+(m_1+m_2)^2][p^2+(m_1-m_2)^2]}      \nonumber \\ 
   && {\kern-120pt} - \frac{1}{8} \left( \frac{1}{p^2+(m_1+m_2)^2} 
      + \frac{1}{p^2+(m_1-m_2)^2} \right)       \ . 
\labbel{43} \end{eqnarray} 
\par 
Eq.(\ref{43}) can be used, in the same way as Eq.(\ref{15}), for obtaining 
the coefficients of the expansions of \( \Sbm \) around particular values 
of \( p^2 \) or a quadrature formula. \par 
The expansion at \( p^2 = 0 \) is straightforward. By writing 
\begin{equation} 
   \Sbm = \overline{S}_1(m_1^2,m_2^2) p^2 + {\cal O}((p^2)^2) \ , 
\labbel{43a} \end{equation} 
Eq.(\ref{43}) gives 
\begin{equation} 
   \overline{S}_1(m_1^2,m_2^2) = 
      \frac{m_1^2m_2^2}{2(m_1^2-m_2^2)^3}\ln\left(\frac{m_1}{m_2}\right) 
      - \frac{m_1^2+m_2^2}{8(m_1^2-m_2^2)^2} \ , 
\labbel{43aa} \end{equation} 
which is just the \( n\to 4 \) limit of \( S_1(n,m_1^2,m_2^2) \), 
Eq.(\ref{19}). 
\par 
More interesting is the \( n\to 4 \) limit of the \( p^2 \to \infty \) 
expansion. Let us start with Eq.(\ref{23}); for \( n\to 4 \) one has 
\begin{equation} 
   (p^2)^{(n-4)/2} \verso{4} 1 + \frac{n-4}{2}\ \ln{p^2} \ 
\labbel{44} \end{equation} 
(see the comments to \( \ln{m_1} \) in Eq.(\ref{39d}) for the appearance 
of \( \ln{p^2}\)). 
As it is known that one loop amplitudes have at most simple poles in 
\( (n-4) \), the leading coefficient \( \Phi^{(s)}_0(n) \), Eq.(\ref{28}), 
can be expanded as 
\begin{equation} 
   \Phi^{(s)}_0(n) \verso{4} \frac{C(n)}{n-4} \phi_0 + \phi_1 \ , 
\labbel{45} \end{equation} 
where \( C(n) \) is again given by Eq.s(\ref{08},\ref{09}) and \( \phi_0 \), 
\( \phi_1 \) are as yet unknown constants. In the \( n\to 4 \) limit the 
large \( p^2 \) expansion of Eq.(\ref{23}) then reads \\ 
\vbox{ \begin{eqnarray} 
   \Sm {\kern+10pt} \verso{4} && 
  \frac{C(n)}{n-4}\phi_0  + \phi_1 + \frac{1}{2} \phi_0 \ln{p^2} \nonumber \\ 
     && {\kern-120pt} + \left[ (m_1^2+m_2^2)\left( 
         \frac{C(n)}{n-4} \frac{2\phi_0+1}{2} 
         + \phi_0 + \phi_1 - \frac{1}{4} + \frac{1}{2}\phi_0 \ln{p^2} \right) 
         \right.      \nonumber \\ && {\kern-100pt} \left. 
     +\frac{1}{2}\left( m_1^2\ln{m_1} + m_2^2\ln{m_2} \right) 
        \phantom { \frac{1}{(p^2)^2} } {\kern-30pt} \right] \frac{1}{p^2} 
     + { \cal O } \left( \frac{1}{(p^2)^2} \right) \ . 
\labbel{46} \end{eqnarray} } 
It is however known from Eq.s(\ref{39a},\ref{39e}) that the pole 
in \( (n-4) \) is independent of \( p^2 \) and of the masses; that implies 
\begin{equation} \phi_0 = -\frac{1}{2} \ , 
\labbel{46a} \end{equation} 
so that Eq.(\ref{46}) becomes 
\begin{eqnarray} 
   \Sm {\kern+10pt} {\buildrel {n \to 4} \over{\longrightarrow}} &&
  - \frac{C(n)}{2(n-4)} + \phi_1 - \frac{1}{4} \ln{p^2} \nonumber \\ 
     && {\kern-120pt} + \left[ 
        \left( \phi_1 - \frac{3}{4} \right) (m_1^2+m_2^2) 
         - \frac{1}{4} m_1^2 \ln{\left(\frac{p^2}{m_1^2}\right)} 
         - \frac{1}{4} m_2^2 \ln{\left(\frac{p^2}{m_2^2}\right)} 
                       \right] \frac{1}{p^2} 
     + { \cal O } \left( \frac{1}{(p^2)^2} \right) \ , 
\labbel{47} \end{eqnarray} 
where the constant \( \phi_1 \) is still unknown. \par 
\newcommand{\sigb}{\overline{\sigma}(m_1^2,m_2^2,p^2)} 
Let us now look at the quadrature formula. We have already seen that the 
replacement Eq.(\ref{30}) transforms Eq.(\ref{15}) into the quadrature 
formula Eq.(\ref{32}). One can similarly put 
\begin{equation} 
   \Sbm = \frac{1}{p^2}\sqrt{ \left(p^2+(m_1+m_2)^2\right) 
                             \left(p^2+(m_1-m_2)^2\right) } \cdot \sigb \ ; 
\labbel{48} \end{equation} 
according to that definition, \( \overline{\sigma}(m_1^2,m_2^2,0) = 0 \). 
Eq.(\ref{43}) then becomes the quadrature formula 
\begin{eqnarray} 
  \sigb &=& \int\limits_0^{p^2} 
  \frac{dx}{\left[ \left(x+(m_1+m_2)^2\right) 
                   \left(x+(m_1-m_2)^2\right) 
           \right]^\frac{3}{2} } \nonumber \\ 
        && \cdot \left[ -\frac{1}{4}x(x+m_1^2+m_2^2) 
                        + \frac{m_1^2m_2^2}{m_1^2-m_2^2} 
                          \ln\left(\frac{m_1}{m_2}\right) \right] \ . 
\labbel{49} \end{eqnarray} 
An elementary integration gives the known \cite{BSH} explicit analytic 
expression 
\begin{eqnarray} 
  \Sbm &=& \frac{1}{4} + \frac{m_1^2+m_2^2}{4(m_1^2-m_2^2)} 
                         \ln\left(\frac{m_1}{m_2}\right) \nonumber \\  
       && {\kern-90pt} + \frac{1}{4p^2} \left[ 
       \sqrt{ \left(p^2+(m_1+m_2)^2\right) \left(p^2+(m_1-m_2)^2\right) } 
       \ln{u(p^2)} +(m_1^2-m_2^2)\ln\left(\frac{m_1}{m_2}\right) \right] \ , 
\labbel{50} \end{eqnarray} 
where 
\begin{equation} 
  u(p^2) = \frac{ \sqrt{p^2+(m_1+m_2)^2} - \sqrt{p^2+(m_1-m_2)^2} } 
                  { \sqrt{p^2+(m_1+m_2)^2} + \sqrt{p^2+(m_1-m_2)^2} } \ . 
\labbel{51} \end{equation} 
\newcommand{\e}{{\mathrm{e}}} 
For real and positive \( p^2 \), \( u(p^2) \) is also real and positive, 
ranging from \( u(+\infty) = 0 \) to \hbox{\( u(0) = m_2/m_1 \)} 
(if \(m_2 < m_1 \)). 
For real and negative \( p^2 > -(m_1-m_2)^2 \), \( u(p^2) \) remains real 
and positive, with \( u(-(m_1-m_2)^2) = 1 \); 
at \( p^2 = -(m_1-m_2)^2 \), further, Eq.(\ref{50}) reads 
\begin{equation} 
  \overline{S}(m_1^2,m_2^2,-(m_1-m_2)^2) = \frac{1}{4} 
       + \frac{m_1m_2}{2(m_1^2-m_2^2)} \ln\left(\frac{m_1}{m_2}\right) \ . 
\labbel{51a} \end{equation} 
From Eq.(\ref{19a}) and Eq.(\ref{17}) a simple explicit calculation gives 
\begin{equation} 
  \lim_{n\to4} \left( S(n,m_1^2,m_2^2,-(m_1-m_2)^2) - 
      S(n,m_1^2,m_2^2,0) \right) = \frac{1}{4} 
       + \frac{m_1m_2}{2(m_1^2-m_2^2)} \ln\left(\frac{m_1}{m_2}\right) \ , 
\labbel{51b} \end{equation} 
so that Eq.(\ref{41}) is satisfied at \( p^2 = -(m_1-m_2)^2 \). \par 
When \( p^2 \) varies from the pseudo-threshold \( -(m_1-m_2)^2 \) to the 
threshold \( -(m_1+m_2)^2 \), one can put 
\begin{equation} 
  \sqrt{p^2+(m_1-m_2)^2} = i \sqrt{-(p^2+(m_1-m_2)^2)} \ ; 
\labbel{52a} \end{equation} 
\( u(p^2) \) then varies on the complex unit circle, 
\( u(p^2) = \e^{i\phi(p^2)} \), with \hbox{\( \phi(-(m_1-m_2)^2) = 0 \) }  
and \( \phi(-(m_1+m_2)^2) = -\pi \), so that in Eq.(\ref{50}) 
 \( \ln{u(p^2)} \) becomes \( i \), the imaginary unit, times an arctangent, 
the square root in front of it is also imaginary and \( \Sbm \) 
remains real (as \( \Sbm \) is analytic at the pseudo-threshold, changing 
\( i \) into \( -i \) in Eq.(\ref{52a}) does not change its value). 
For \( p^2 < -(m_1+m_2)^2 \), finally, \( u(p^2) \) is 
negative, ranging from \( -1 \) at \( p^2 = -(m_1+m_2)^2 \) 
to \( 0 \) at \( p^2 = - \infty \); correspondingly, the square 
root of Eq.(\ref{50}) is again real, while \( \ln{u(p^2)} \) becomes complex, 
acquiring an imaginary part \( i\pi \) whose sign however can be determined 
only by giving an infinitesimal imaginary part to \( p^2 \) (if 
\( \mathrm{Im} p^2 < 0 \), the imaginary part is \( i\pi \)). \par 
\par 
Further, on account of the explicit analytic expression Eq.(\ref{50}), 
in the \( p^2 \to \infty \) limit Eq.(\ref{41a}) becomes 
\begin{eqnarray} 
   \Sm {\kern+10pt} {\buildrel {n \to 4} \over{\longrightarrow}} &&
  - \frac{C(n)}{2(n-4)} + \frac{1}{2} - \frac{1}{4} \ln{p^2} \nonumber \\ 
     && {\kern-120pt} - \frac{1}{4} \left[ 
          m_1^2+m_2^2 
         + m_1^2 \ln{\left(\frac{p^2}{m_1^2}\right)} 
         + m_2^2 \ln{\left(\frac{p^2}{m_2^2}\right)} 
                       \right] \frac{1}{p^2} 
     + { \cal O } \left( \frac{1}{(p^2)^2} \right) \ ; 
\labbel{52} \end{eqnarray} 
the comparison with Eq.(\ref{47}) gives 
\begin{equation} \phi_1 = \frac{1}{2} \ . 
\labbel{53} \end{equation} 
Eq.s(\ref{46a},\ref{53}) can be obtained also by the explicit integration 
of Eq.(\ref{35}) in the \( n \to 4 \) limit. 
\par 
Similarly, one can expand Eq.(\ref{50}) around the physical threshold 
\( p^2 = - (m_1+m_2)^2 \), obtaining, in the notation of Eq.(\ref{29d}), 
\begin{equation} \Psi_0^{(s)}(4) = - \frac{\pi}{4} \ . 
\labbel{54} \end{equation} 
\section{Outlook. } \par 
As already said in the Introduction, the 1-loop self-mass is too simple 
for really showing the potential of the new method. It is nevertheless 
possible to try to assess which features of the method can be reasonably 
expected to apply also to less elementary cases. \par 
To start with, it is apparent from the derivation that a linear system 
of first order differential equations can be established, in full 
generality, for multi-point and multi-loop graphs as well, as the 
integration by part identities can be written for any number of external 
lines and loops. \par 
In the case of more general multi-point amplitudes, depending on a set of 
independent external vectors \( p_{i,\mu} \), one has simply to generalize 
Eq.(\ref{14}) by considering all the various 
\( p_{i,\mu} \frac{\partial}{\partial p_{j,\mu} } \) combinations and then 
work out some kinematical algebra for extracting the derivative with respect 
to any given external scalar variables. A first application to the 1-loop 
vertex is underway \cite{CR}. \par 
In the general multi-loop case several independent master integrals are 
expect to appear, so that rather than a single first order equation 
there will be a linear system of first order differential equations, whose 
coefficients however will be in full generality ratios of suitable polynomials 
in the dimension \( n \), the masses and the external scalar variables 
(work on the 2-loop ``sunrise" self-mass graph is in progress \cite{CCR}). 
In the general case, the equations will involve, besides the master integrals 
for the considered original Feynman graph, a number of ``inhomogeneous terms" 
corresponding to the master integrals for all the related, simpler Feynman 
graphs (with the same number of loops) obtained by removing a denominator from 
the original graph; those simpler graph amplitudes are to be considered 
known --- or else can be studied recursively by the same algorithm. \par 
In the general multi-point, multi-loop case and for arbitrary values of 
the masses simple quadrature formulae and closed analytic results 
in terms of known functions cannot be expected to exist, but the equations 
will be anyhow a powerful tool for studying the value of the amplitudes for 
particular values of the occurring masses as well as for attempting, in 
particularly simple cases, the analytic integrations. \par 
The equations are 
particularly well suited for the expansion in \( (n-4) \); as the equations 
are identities in \( n \), the \( n\to4 \) limit can be worked out at once 
for the whole multi-loop amplitudes, without any reference to the 
\( n\to4 \) limit of the inserted loops, so that any overlapping divergencies 
problem is avoided. \par 
Once a system of equations for quantities finite in the \( n\to4 \) limit 
has been obtained, its numerical integration is straightforward, for 
virtually any number of loops and external lines, provided suitable starting 
points are given. It is reasonable to think (but also confirmed by some 
preliminary work, \cite{CCR}) that the absence of the would-be kinematical 
singularities, such as those at the \( p^2 = 0 \) for the self-mass, will 
provide with useful information for the evaluation of those starting points. 
\par 
The equations will also be very useful, in general, to provide any kind of 
required expansion of the graph amplitudes, for instance for asymptotically 
large values of the 
scalar variables or at the physical thresholds, in terms of very few 
constants; even if those constants are not expected to be fixed by the 
equations themselves, the information which they contain can be exploited 
to make easier the calculation of those missing values. 
\vskip 1 truecm 
\noindent 
{\bf Acknowledgments.} The author is pleased to thank M. Caffo, H. Czyz 
and S. Laporta for several clarifying discussions. 
\def\NP{{\sl Nuc. Phys.}} 
\def\PL{{\sl Phys. Lett.}} 
\def\PR{{\sl Phys. Rev.}} 
\def\PRL{{\sl Phys. Rev. Lett.}} 

\end{document}